\documentclass[%
prl,twocolumn,10pt,aps,longbibliography]{revtex4-1}

\usepackage{graphicx}
\usepackage{dcolumn}
\usepackage{bm}
\usepackage[caption=false]{subfig}
\usepackage{amssymb}
\usepackage{amsmath}

\begin{document}
	
	\preprint{INTEL-CONFIDENTIAL}
	
	\title{Inversion Charge-boost and Transient Steep-slope induced by Free-charge-polarization Mismatch in a Ferroelectric-metal-oxide-semiconductor Capacitor}
	
	\author{Sou-Chi Chang}
	\email{sou-chi.chang@intel.com}
	\affiliation{Components Research, Intel Corporation, Hillsboro, Oregon 97124, USA}
	
	\author{Uygar E. Avci}%
	\affiliation{Components Research, Intel Corporation, Hillsboro, Oregon 97124, USA}
	
	\author{Dmitri E. Nikonov}%
	\affiliation{Components Research, Intel Corporation, Hillsboro, Oregon 97124, USA}
	
	\author{Ian A. Young}%
	\affiliation{Components Research, Intel Corporation, Hillsboro, Oregon 97124, USA}

	\begin{abstract}
		In this letter, the transient behavior of a ferroelectric (FE) metal-oxide-semiconductor (MOS) capacitor is theoretically investigated with a series resistor. It is shown that compared to a conventional high-k dielectric MOS capacitor, a significant inversion charge-boost can be achieved by a FE MOS capacitor due to a steep transient subthreshold swing (SS) driven by the free charge-polarization mismatch. It is also shown that the observation of steep transient SS significantly depends on the viscosity coefficient under Landau's mean field theory, in general representing the average FE time response associated with domain nucleation and propagation. Therefore, this letter not only establishes a theoretical framework that describes the physical origin behind the inversion charge-boost in a FE MOS capacitor, but also shows that the key feature of depolarization effect on a FE MOS capacitor should be the inversion-charge boost, rather than the steep SS (e.g., sub-60mV/dec at room temperature), which cannot be experimentally observed as the measurement time is much longer than the FE response. Finally, we outlines the required material targets for the FE response in field-effect transistors to be applicable for next-generation high-speed and low-power digital switches.
	\end{abstract}
	
	\keywords{Ferroelectric Transistor, Polarization, Negative Capacitance}
	
	\maketitle
\begin{figure}
	\includegraphics[width=0.7\linewidth]{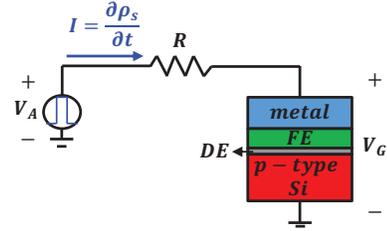}
		\caption{The schematic of a $R-C_{FEMOS}$ circuit studied in this work, where $C_{FEMOS}$ is a FE MOS capacitor with $p$-type silicon. $V_{A}$ is the applied voltage, $V_{G}$ is the gate voltage across the capacitor, $R$ is the resistance of series resistor, and current flowing through the resistor, $I$, is defined as $\frac{\partial \rho_{s}}{\partial t}$ with $\rho_{s}$ being the free charge density provided by the external circuitry.}
	\label{fig1}
	\end{figure}

	The relentless pursuit of Moore's law in the past four decades leads to a significant improvement in the computing power of modern microprocessors \cite{658762}. However, as the scaling of CMOS transistors continues, the on-off current ratio is dramatically reduced, and thus the increasing static power makes the circuit design more and more difficult for high energy-efficient applications \cite{1250885}. In 2008, a transistor structure using FE materials in the gate stack was proposed to improve SS through stabilizing FE negative capacitance in the steady state \cite{doi:10.1021/nl071804g}. However, the central idea of this proposal is all based on the $S$-shape of polarization-electric field relation given by Landau's free energy functional, which has been under debate if it can physically describe the polarization switching in the FE oxide \cite{doi:10.1021/nl1037215}. In 2015, a transient differential negative capacitance was reported from a circuit composed of a resistor and a FE capacitor in series \cite{Khan2015}. Our recent work theoretically proved that this differential negative capacitance is driven by the free charge-polarization mismatch as qualitatively pointed out in Ref. \cite{Catalan2015}, which can be well described under Landau's mean field theory \cite{PhysRevApplied.9.014010}. In 2018, the transient differential negative capacitance has been also experimentally observed in a FE MOS capacitor \cite{8186219}, indicating the existence of transient mismatch between bound charge and free charge in a MOS structure. The effect of free charge-polarization mismatch on transient behavior of a FE transistor was qualitatively discussed in Ref. \cite{NG201712}, where the importance of measurement time in FE transistor characterization is emphasized but no inversion charge-boost in the steady-state from a FE transistor is concluded. Hence, in this letter, a numerical model is developed to investigate the transient charging behavior of a FE MOS capacitor in the R-C circuit as shown in Fig. \ref{fig1}. From the model, it is found that an inversion charge-boost, leading to an increase in the drive-current of a transistor, can be achieved by transient SS driven by a free charge-polarization mismatch. The impact of FE response time on the charge-boost are also investigated, and therefore the material target for a FE field-effect transistor to be applicable for the next generation digital switch are identified.
	
	To describe the transient response of a $R-C_{FEMOS}$ circuit shown in Fig. \ref{fig1}, similar to the approach given in Ref. \cite{PhysRevApplied.9.014010}, Kirchhoff's law is used to describe the free charge density flowing through the resistor given as
	\begin{eqnarray}
	\frac{\partial \rho_{s}}{\partial t}=\frac{V_{A}-V_{G}}{AR},
	\label{eq1}
	\end{eqnarray}
	where $\rho_{s}$ is the free charge density on a capacitor, $V_{A}$ is the applied voltage, $V_{G}$ is the gate voltage, $A$ is the cross-sectional area of the capacitor, and $R$ is the resistance of a series resistor. The average polarization dynamics of the FE oxide, $P$, is based on Landau's mean field theory given as \cite{PhysRevB.20.1065,PhysRevB.88.024106,PhysRevApplied.4.044014,PhysRevB.68.094113}
	\begin{eqnarray}
	\gamma\frac{\partial P}{\partial t}&=&-\frac{\partial U}{\partial P} \nonumber \\
	&=&-2\alpha_{1}P-4\alpha_{11}P^{3}-\alpha_{111}P^{5}+E_{FE},
	\label{eq2}
	\end{eqnarray}
	where $\alpha_{1}$, $\alpha_{11}$, and $\alpha_{111}$ are Landau expansion coefficients describing a double-well FE free energy profile, $U$, and $E_{FE}$ is the electric field across the FE oxide. From electrostatics, the voltage across the FE oxide, $V_{FE}$, is given as
	\begin{eqnarray}
	V_{FE} = E_{FE}t_{FE} = \frac{t_{FE}\left(\rho_{s}-P\right)}{\epsilon_{0}},
	\label{eq3}
	\end{eqnarray}
	in which $t_{FE}$ is the FE oxide thickness and $\epsilon_{0}$ is the vacuum dielectric constant. In a FE MOS capacitor, the gate voltage has to be shared within the vertical stack; namely, 
	\begin{eqnarray}
	V_{G} = V_{metal} + V_{FE} + V_{DE} + V_{si} + V_{fb},
	\label{eq4}
	\end{eqnarray}
	, where $V_{metal}$ and $V_{DE}$ are the voltages across metal and the dielectric (DE), respectively. $V_{si}$ is the silicon surface potential drop and $V_{fb}$ is the flat-band voltage given as $\phi_{1} -\phi_{2} - E_{f,metal} - \delta$ with $\phi_{1}$ and $\phi_{2}$ being conduction band discontinuities at the metal/FE and DE/silicon interfaces, respectively, $E_{f,metal}$ is the Fermi energy of metal, and $\delta$ is the energy difference between conduction band and Fermi level in the quasi-equilibrium region of $p$-type silicon. For a given $\rho_{s}$, $V_{metal}$ and $V_{DE}$ are given as \cite{PhysRevApplied.7.024005}
	\begin{eqnarray}
	V_{metal} &=& \frac{\rho_{s}\lambda_{metal}}{\epsilon_{metal}\epsilon_{0}}, \label{eq5} \\
	V_{DE} &=& \frac{\rho_{s}t_{DE}}{\epsilon_{DE}\epsilon_{0}},
	\label{eq6}
	\end{eqnarray}
	where $\lambda_{metal}$ and $\epsilon_{metal}$ are the screening length and relative dielectric constant of metal, respectively, and $t_{DE}$ and $\epsilon_{DE}$ are the thickness and relative dielectric constant of DE, respectively. With a given $\rho_{s}$, the induced net charge at the silicon side, $Q_{net}=-\rho_{s}$, and corresponding $V_{si}$ are obtained by self-consistently solving Shr\"{o}dinger and Poisson equations given from Eqs. \ref{eq11} to \ref{eq14} with Newton-Raphson method for convergence \cite{doi:10.1063/1.346245}, where $\left(100\right)$ wafer orientation is used for silicon.
	\begin{widetext}
	\begin{eqnarray}
	\left[-\frac{\hbar^{2}}{2m_{j}}\frac{\partial^{2}}{\partial x^{2}}\pm eV_{si}\left(x\right)\right]\psi_{ij}\left(x\right)&=&E_{ij}\psi_{ij}\left(x\right), \label{eq11} \\
	n\left(x\right)&=&\sum_{j=l,t}\frac{k_{B}Tg_{j}m_{d,j}}{\pi \hbar^{2}}\sum_{i}\log\left[1+e^{\frac{E_{f,si}-E_{ij}}{k_{B}T}}\right]|\psi_{ij}\left(x\right)|^{2}, \label{eq12} \\
	p\left(x\right)&=&\sum_{j=hh,lh}\frac{k_{B}Tm_{d,j}}{\pi \hbar^{2}}\sum_{i}\log\left[1+e^{\frac{E_{f,si}-E_{ij}}{k_{B}T}}\right]|\psi_{ij}\left(x\right)|^{2}\\ \nonumber &+&\frac{k_{B}Tm_{d,so}}{\pi \hbar^{2}}\sum_{i}\log\left[1+e^{\frac{E_{f,si}-E_{i,so}-\Delta_{so}}{k_{B}T}}\right]|\psi_{i,so}\left(x\right)|^{2}, \label{eq13} \\
	\frac{\partial^{2} V_{si}\left(x\right)}{\partial x^{2}}&=&\frac{-Q_{net}}{\epsilon_{si}\epsilon_{0}}=\frac{-e}{\epsilon_{Si}\epsilon_{0}}\left[p\left(x\right)-n\left(x\right)-N_{A}\right], \label{eq14}
	\end{eqnarray}
	\end{widetext}
	in which $m_{j}$ is the effective mass of valley $j$ for electrons or holes, $e$ is the elementary charge, $\psi_{ij}$ and $E_{ij}$ are the envelop function and energy of valley $j$ in the subband $i$, respectively, $k_{B}$ is the Boltzmann constant, $T$ is the temperature, $g_{i}$ is the valley degeneracy factor, $m_{d,j}$ is the density-of-state effective mass of valley $j$ for electrons or holes, $\hbar$ is the reduced Planck constant, $E_{f,si}$ and $V_{si}\left(x\right)$ are the Fermi level in the quasi-equilibrium region and the surface potential profile within $p$-type silicon, respectively, $\epsilon_{si}$ is the relative dielectric constant of silicon, and $N_{A}$ is the $p$-doped (acceptor) concentration. Eqs. \ref{eq1} to \ref{eq14} are solved numerically for convergence at every time step to describe the transient charging and discharge behavior of a FE MOS capacitor in series with a resistor. All the simulation parameters are listed in the supplementary material. Note that in the following simulation results, high-k MOS capacitors are also shown to help illustrate the concept, and the only modification in the model is replacing $V_{FE}$ in Eq. \ref{eq3} with the voltage across the high-k dielectric, $V_{high-k}$, given as
	\begin{eqnarray}
	V_{high-k} &=& \frac{\rho_{s}t_{high-k}}{\epsilon_{high-k}\epsilon_{0}},
	\label{eq7}
	\end{eqnarray}
	where $t_{high-k}$ and $\epsilon_{high-k}$ are the thickness and relative dielectric constant of the high-k layer. The numerical model described above is justified by qualitatively capturing the transient negative differential capacitance reported in the recent experimental measurements when a bipolar voltage pulse is applied to a hafnium-based FE MOS capacitor \cite{8186219}, as given in the supplementary material.
	
	\begin{figure}
		\includegraphics[width=1.05\linewidth]{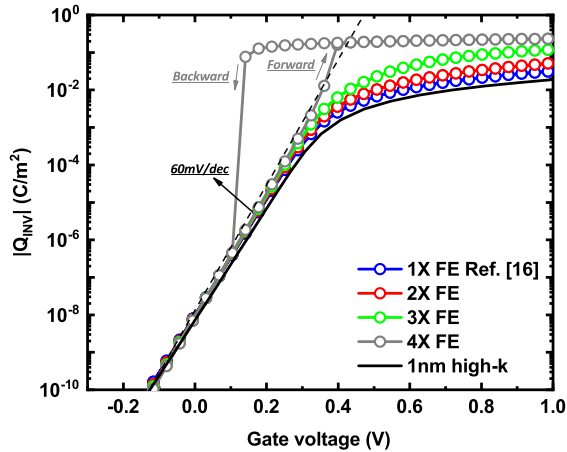}
		\caption{Inversion charge density at room temperature as a function of gate voltage with different FE strengths and the high-k dielectric under steady state. $60$mV/dec is given as a black dash line for reference. The directions of forward and backward sweep are also identified. Parameters for $1X$, $2X$, $3$, and $4X$ FE are defined in the supplementary material, where the parameters for $1X$ FE are extracted by fitting the experimental measurements in Ref. \cite{7838402}.}
		\label{fig3}
	\end{figure}
	
	Figure \ref{fig3} shows the inversion charge density under forward and reverse applied voltage sweeps, where the pulse duration for each applied voltage is long enough to make sure the charge on a capacitor in the RC circuit reaches the steady state. This scenario can be considered as a slow DC sweep in the experimental measurements as long as the FE response is not too slow. From Fig. \ref{fig3}, it can be seen that a significant charge boost can be achieved in the strong inversion region in a FE MOS capacitor compared to the conventional high-k one. A stronger FE oxide delivers a greater charge boost due to a larger depolarization field across the FE oxide, which will be explained in details below. Note that a stronger FE oxide here means a wider double-well free energy profile with a deeper barrier height, which physically corresponds to a larger coercive field and also a greater remanent polarization. This trend is also consistent to the result based on the classical approach to the relation between inversion charge and silicon surface potential \cite{8027211}. Interestingly, Fig. \ref{fig3} implies that when the measurement time is much slower than the FE response, a significant charge-boost can be delivered without SS sharper than $60$mV/dec in the forward sweep.
	
	\begin{figure}
	\subfloat[]{%
		\includegraphics[width=1.05\linewidth]{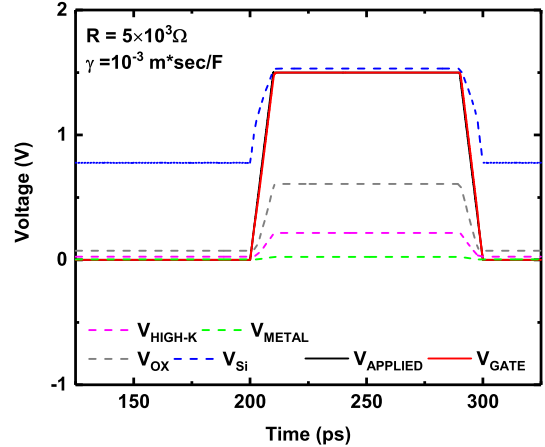}%
	} \\
	\subfloat[]{%
		\includegraphics[width=1.05\linewidth]{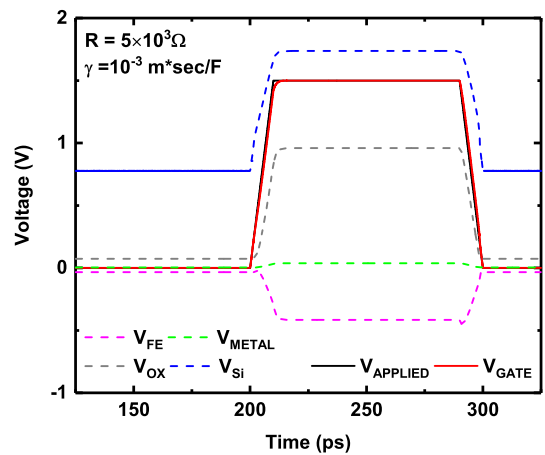}%
	}
	\caption{Transient voltage distribution in each layer of (a) high-k and (b) FE MOS capacitors when a positive voltage pulse is applied.}
	\label{fig4}
	\end{figure}
	
	Next, the underlying mechanism that drives this inversion charge-boost in the steady state is investigated by applying a voltage pulse with a fixed magnitude to a FE MOS capacitor. Figures \ref{fig4}(a) and (b) show the transient voltage response in each layer for both high-k and FE MOS capacitors, respectively, and it can be immediately observed that the main difference between high-k and FE capacitors is the opposite sign of voltage drops across the high-k and FE layers in the steady state. Also, under the steady state, one can see that the direction of field across the FE oxide in the case of charge boost is opposite to that of polarization, which is the main feature of depolarization \cite{PhysRevApplied.7.024005}. Hence, to achieve the inversion charge boost in the steady state, the field across the FE oxide has to be dominated by the depolarization, rather than the applied bias, which is also consistent to the thermodynamic picture given in Ref. \cite{8027211}. More importantly, from Fig. \ref{fig4}(b), to make the depolarization dominate the field across the FE oxide in the steady state, it is necessary for a capacitor to have a transient region where the voltage drop across the FE oxide decreases with time; that is, $\frac{\partial V_{FE}}{\partial t}<0$. This negative $\frac{\partial V_{FE}}{\partial t}$ leads to a larger $\frac{\partial V_{si}}{\partial t}$ (or transient steep SS, $\frac{\partial V_{si}}{\partial V_{g}}$) and thus gives a FE MOS capacitor a faster increase in the inversion charge.
	
	\begin{figure}
		\subfloat[]{%
			\includegraphics[width=1.05\linewidth]{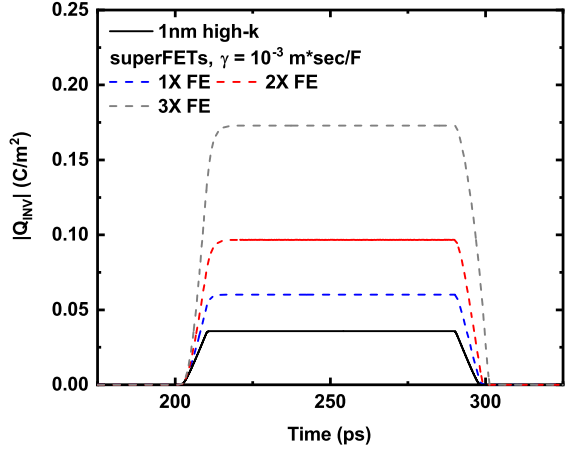}%
		} \\
		\subfloat[]{%
			\includegraphics[width=1.05\linewidth]{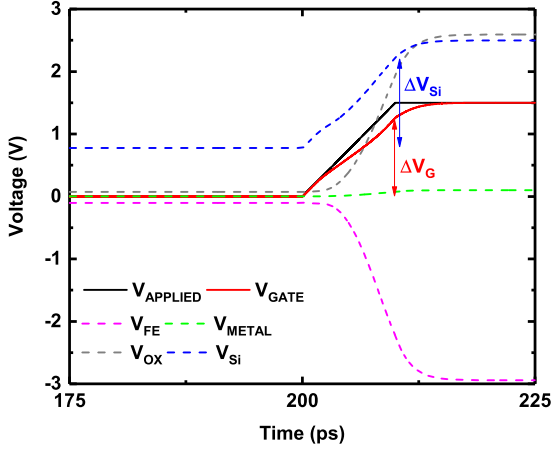}%
		}
		\caption{(a) Inversion charge density as a function of time when applying a positive pulse to a FE MOS capacitor with different ferroelectric strengths of oxide. The inversion charge of a high-k dielectric MOS capacitor is also given for the reference of charge boost in FE ones (the black). (b) Transient voltage distribution as a function of time for the highest charge boost shown in Fig. \ref{fig5}(a).}
		\label{fig5}
	\end{figure}
	
	The inversion charge boost driven by transient steep SS can be seen more clearly from Fig. \ref{fig5}(a), where the boost increases with stronger ferroelectricity. Note that for all the cases shown in Fig. \ref{fig5}(a), none of them shows sub-60mV/dec during the steady-state forward voltage sweep as can be seen in Fig. \ref{fig3}. The field distribution as a function of time in each layer for the largest boost in Fig. \ref{fig5}(a) is given in Fig. \ref{fig5}(b), where $\Delta V_{si}>\Delta V_{G}$ is labeled and the corresponding average SS is $\sim 52$ mV/dec at room temperature (RT) based on $SS=\left(\frac{\partial V_{si}}{\partial V_{G}}\frac{\partial \log_{10}I_{ds}}{\partial V_{si}}\right)^{-1}$ with $I_{ds}$ being the channel current. Therefore, the key feature of depolarization effect on a FE MOS capacitor under DC or slow measurements should be the inversion charge-boost, rather than sharp SS (e.g., sub 60mV/dec at RT), which may only occur in the transient state due to free charge-polarization mismatch as explained below. 
	
	Similar to Ref. \cite{PhysRevApplied.9.014010}, the physical origin of this transient SS can be understood by taking the time derivative of Eq. \ref{eq3} as shown below.
	\begin{eqnarray}
	\frac{\partial V_{FE}}{\partial t} = \frac{t_{FE}}{\epsilon_{0}}\left(\frac{\partial \rho_{s}}{\partial t}-\frac{\partial P}{\partial t}\right).
	\label{eq8}
	\end{eqnarray}
	From Eq. \ref{eq8}, it can be seen that during charging of a FE MOS capacitor, both $\frac{\partial \rho_{s}}{\partial t}$ and $\frac{\partial P}{\partial t}$ are positive; therefore, the only possibility to make $\frac{\partial V_{FE}}{\partial t}$ negative is $\frac{\partial P}{\partial t}>\frac{\partial \rho_{s}}{\partial t}$; that is, the bound charge in the FE oxide changes faster than free charge on a capacitor. Since the capacitance of the FE layer in a FE MOS capacitor, $C_{FE}$, is mathematically defined as
	\begin{eqnarray}
	C_{FE}=\frac{\partial \rho_{s}}{\partial V_{FE}},
	\label{eq9}
	\end{eqnarray}
	$C_{FE}$ transiently becomes negative as the mismatch between free charge and polarization in the ramping rate occurs.
	
\begin{figure}
		\includegraphics[width=1.05\linewidth]{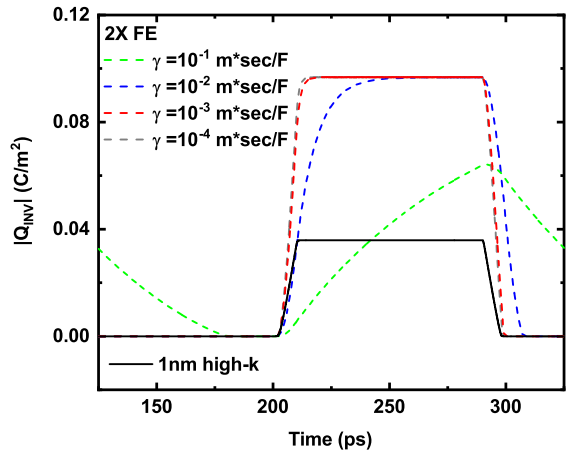}
		\caption{Inversion charge density as a function of time for a FE MOS capacitor with different viscosity coefficients. The one with high-k dielectric is also shown for reference (the black).}
		\label{fig6}
	\end{figure}

	From Eq. \ref{eq8}, it can be seen that the FE response time plays a significant role for transient steep SS and thus the inversion charge boost. Based on Landau theory, Eq. \ref{eq8} can be re-written into
	\begin{eqnarray}
	\frac{\partial V_{FE}}{\partial t} = \frac{t_{FE}}{\epsilon_{0}}\left(\frac{\partial \rho_{s}}{\partial t}+\frac{1}{\gamma}\frac{\partial U}{\partial P}\right).
	\label{eq10}
	\end{eqnarray}
	Eq. \ref{eq10} shows that the reduction of $\gamma$, which means a faster FE response, leads to a more significant transient SS, as can be seen in Fig. \ref{fig6}, where $\gamma$ less than $10^{-3}$ m*sec/F is roughly required for ramping up the inversion charge within $10$ps in a $100\times20$nm$^{2}$ FE MOS capacitor. Note that from Ref. \cite{PhysRevApplied.9.014010}, $\gamma$ extracted from a $50\times50\mu$m$^{2}$ FE doped hafnium capacitor is roughly between $10^{3}$ to $10^{4}$ m*sec/F depending on the amount of FE domains switched throughout the film. Therefore, it is important to characterize the value of $\gamma$ as a capacitor is scaled down or a new FE material is introduced for logic applications, which is beyond the scope of this letter. Moreover, Fig. \ref{fig6} also implies that the characterization time in $I_{ds}$-$V_{G}$ measurements is also critical to observe the charge-boost due to the finite FE response time; that is, to observe a sharp SS in the measurements, the time scale of significant free charge-polarization mismatch has to be in the same order to that of voltage sweeps. 
	
	In conclusion, this letter theoretically shows that the boost in inversion charge density in a FE MOS capacitor (compared to the high-k one) can be achieved by transient SS driven by polarization-free charge mismatch during charging and discharge of a capacitor. It is also shown that capacitance associated with the FE layer mathematically becomes negative as the polarization changes faster than the free charge on a FE MOS capacitor. Since the transient SS highly depends on how fast the FE polarization can respond to an electric field, the material targets for the FE response time are also identified for a FE field-effect-transistor to be suitable for the next-generation digital switch.
	
	\bibliography{apssamp}
	
\end{document}